\documentclass[reprint,showpacs,twocolumn,superscriptaddress,aps,pra]{revtex4-2}

\usepackage[utf8]{inputenc}
\usepackage[american,british]{babel}
\usepackage[T1]{fontenc}
\usepackage[pdftex]{graphicx}  
\usepackage{graphicx, xcolor}
\usepackage{dcolumn}
\usepackage{physics}

\usepackage{bm}
\usepackage{amsmath,amsthm,amssymb}
\usepackage{color}
\usepackage{verbatim} 
\usepackage{soul}

\usepackage{multirow}

\usepackage[normalem]{ulem}
\usepackage{natbib}

\newcommand{\smt}{Science, Mathematics and Technology Cluster, Singapore University of Technology and Design, 8 Somapah Road, 487372 Singapore}
\newcommand{\epd}{Engineering Product Development Pillar, Singapore University of Technology and Design, 8 Somapah Road, 487372 Singapore}
\newcommand{\cqt}{Centre for Quantum Technologies, National University of Singapore 117543, Singapore}
\newcommand{\majulab}{MajuLab, CNRS-UNS-NUS-NTU International Joint Research Unit, UMI 3654, Singapore}

\definecolor{darkGreen}{RGB}{0,110,0}
\definecolor{darkBlue}{RGB}{0,0,130}
\usepackage[hidelinks]{hyperref}

\usepackage[normalem]{ulem}

\hypersetup{
 colorlinks=true,
 linkcolor=blue,
 citecolor = magenta,
 urlcolor = blue,
pdfborder={0 0 0},
}

\usepackage{dsfont}

\begin{document}
\preprint{APS/123-QED}
 
\title{Exploring the performance of superposition of product states: \\ from 1D to 3D quantum spin systems}

\author{Apimuk Sornsaeng}
\affiliation{\smt}
\affiliation{\cqt} 

\author{Itai Arad}
\affiliation{\cqt} 

\author{Dario Poletti}
\email{dario\_poletti@sutd.edu.sg}
\affiliation{\smt} 
\affiliation{\cqt} 
\affiliation{\epd} 
\affiliation{\majulab}

\date{\today}
\begin{abstract}
  Tensor networks (TNs) are one of the best available tools to study
  many-body quantum systems. TNs are particularly suitable for
  one-dimensional local Hamiltonians, while their performance for
  generic geometries is mainly limited by two aspects: the
  limitation in expressive power and the approximate extraction of
  information.  Here we investigate the performance of
  \textit{superposition-of-product-states} (SPS) ansatz, a
  variational framework structurally related to canonical polyadic
  tensor decomposition. The ansatz does not compress information as
  effectively as tensor networks, but it has the
  advantages (i) of allowing accurate extraction of information,
  (ii) of being structurally independent of the geometry of the
  system, (iii) of being readily parallelizable, and (iv) of
  allowing analytical shortcuts.  We first study the typical
  properties of the SPS ansatz for spin$-1/2$ systems, including its
  entanglement entropy, and its trainability.  We then use this
  ansatz for ground state search in tilted Ising models---including
  one-dimensional and three-dimensional with short- and long-range interaction, and a
  random network---demonstrating that SPS can attain high accuracy. 
\end{abstract}

\maketitle

\section{Introduction}

The primary difficulty in studying quantum
many-body physics lies in the exponential growth of the Hilbert
space as the number of particles increases \cite{landau2015physics,nielsen2010quantum}. Various
numerical and analytical tools have been developed to capture the
ground state properties of quantum many-body systems.  A prominent
and very effective class of numerical methods is variational
algorithms, which include, for example, variational Quantum Monte
Carlo \cite{foulkes2001quantum}, Neural Quantum States \cite{CarleoTroyer,variational_benchmark}, and Tensor Network
approaches \cite{schollwock2011density,orus2014practical,verstraete2004renormalization}. Within the variational methods, we can list two
main categories: those that sum over all the configurations, and
those that use sampling. Variational Quantum Monte Carlo and Neural
Quantum States belong to the latter category while tensor network
algorithms typically belong to the first one. In tensor networks
algorithms, full summation over all configurations (i.e., exact
contraction of the network), is often infeasible, as it may require
exponential computational cost. One is therefore led to using
approximate tensor network contractions, such as Corner Transfer
Matrix Renormalization Group (CTMRG) \cite{nishino1996corner,verstraete2004renormalization,orus2009simulation}
boundary MPS \cite{jordan2008classical,weyrauch2013efficient}, and belief propagation to name a few
\cite{robeva2019duality,alkabetz2021tensor,sahu2022efficient,guo2023block}. However, in some of the most
challenging simulations, even tensor network algorithms resort to
sampling techniques \cite{sandvik2007variational,wang2011monte,ferris2012perfect}. 

When using these methods, errors can creep in for different reasons.
For instance, when one performs samples, the expectation values may
not be accurately estimated, leading to a sampling error. When using
tensor networks beyond one dimension, while one can sum over all
configurations, the sum is evaluated in an approximated manner,
which also can lead to errors.

In tensor networks, approximations stem either from reduced expressibility of the ansatz, or from the complexity of accurately extracting information from it. 
Here, we aim to use an ansatz which can readily be extended to multiple dimensions and from which one can readily and accurately extract information. Furthermore, the computations with this ansatz are highly parallelizable, leading to potential speed up.

From a physical point of view, the ansatz is a \emph{Superposition
of Product States} (SPS), which can be viewed as an extension of the
bosonic Gutzwiller ansatz \cite{RoksharKotliar1991, KrauthBouchaud1992, JakschZoller1998} to a superposition of non-orthogonal
product states. From a more mathematical
perspective, the SPS ansatz can be understood as a rank-$M$
canonical polyadic (CP) decomposition of the tensor that is defined
by coefficients of the underlying quantum state in the standard basis. This
decomposition is also known as CANDECOMP/PARAFAC \cite{CP1,CP2,CP3}.
From this perspective, any tensor can be expressed as a sum of $M$
rank-$1$ tensors, each formed by the tensor product of local
vectors, with coefficients that weight the contribution of each
rank-1 tensor. The expressive power of the SPS ansatz is controlled
by the decomposition rank $M$, i.e., the number of product states in
the decomposition.

The structure of the rest of this paper is as follows. In
Sec.~\ref{sec: SPS}, we formally present the construction of the SPS
ansatz and examine its statistical properties, including typicality,
expressive power, and trainability. In Sec.~\ref{sec: GS}, we
benchmark the performance of the SPS ansatz in ground-state search
tasks for various lattice geometries of the tilted Ising model,
including 1D, 3D, long-range systems, and random coupling networks.
A summary and discussion of our findings are given in Sec.~\ref{sec:
conclusion}. For completeness, analytical derivations of the key
quantities employed in our analysis are provided in App.~\ref{appx:
analytic}. We also provide details of the connectivity in the random-coupling system in App.~\ref{app: conn}.

\section{SPS ansatz}\label{sec: SPS}

A superposition-of-product-states (SPS) ansatz for an $L$-sites
quantum system is constructed by linearly combining $M$
\textit{normalized} product states with arbitrary amplitudes $c_m$.
In general, this can be written as 
\begin{equation}
  \ket{\Psi} = \frac{\ket{\Psi_u}}{\sqrt{\mathcal{Z}}}  
    = \frac{1}{\sqrt{\mathcal{Z}}} 
      \sum_{m=1}^M c_m \bigotimes_{l=1}^L \ket{\Psi^m_l}, 
\label{eq:sps_ansatz}
\end{equation}  
with the unnormalized superposition of product states
$\ket{\Psi_u}$, and $\ket{\Psi^m_l}$ is a state defined on a single
site $l$. Since each product state is in general non-orthogonal to
others, we have added the normalization $\mathcal{Z} :=
\braket{\Psi_u}$. Equation~\eqref{eq:sps_ansatz} describes a general
SPS ansatz.

In the following, we will focus on spin-$1/2$ systems whose ground states can be represented with real coefficients in the computational basis. Thus, in this work, we can specifically consider the following SPS
ansatz 
\begin{equation}
  \ket{\Psi^m_l} := \ket{\theta^m_{l}} 
    =\mqty(\cos\theta_l^{m} \\ 
  \sin\theta_l^{m}), \label{eq:single_sps}
\end{equation}  
where all parameters $\Theta := \{c_m,\theta^m_l\}$ are real-valued,
resulting in a total of $(L + 1)M$ parameters.

In the following, we are going to analyze statistical properties of
the initially prepared arbitrary SPS in
Sec.~\ref{ssec:statistics_wave}, particularly its normalization
factor $\mathcal{Z}$ (Sec.~\ref{ssec:statistics_norm}), the local
observables (Sec.~\ref{ssec:statistics_observables}), and
entanglement (Sec.~\ref{ssec:statistics_entanglement}), and then we
further study its trainability in Sec.~\ref{ssec:trainability}.

\subsection{Statistics of the SPS ansatz}
\label{ssec:statistics_wave}

To examine the statistical behavior of the SPS ansatz $\ket{\Psi}$,
we will analyze the expectation value and variance of a
local observable, and further explore
the expressive power of the ansatz via the entanglement entropy
that it can produce. In this section, we
consider a random SPS with $\theta_l^{m}$ initialized uniformly in
$\mathcal{D}_\theta = [-\pi/2, \pi/2]$ and $c_m$ in $\mathcal{D}_c =
[-1,1]$, thus, the SPS parameters $\Theta$ are drawn from
$\mathcal{D} = \mathcal{D}_c^{M} \times \mathcal{D}_\theta^{LM}$.

\subsubsection{Statistics of the norm} 
\label{ssec:statistics_norm}  

To further understand the global structure of the SPS states in
Hilbert space, we first examine its normalization factor
$\mathcal{Z}$ to find how stable the ensemble is under random
sampling. From Eqs.~\eqref{eq:sps_ansatz} and \eqref{eq:single_sps},
we can analytically write the normalization constant as 
\begin{equation*}
  \mathcal{Z} = \sum_{m,m'=1}^M c_mc_{m'}\prod_{l=1}^L 
    \cos(\theta_{l}^{m}-\theta_{l}^{m'}).
\end{equation*}
Using well-known formulas describing the statistics of uniform
distributions, we conclude that the diagonal terms, $m = m'$
contribute $c_m^2$ with $\mathbb{E}_{\mathcal{D}_c}[c^2_m] = 1/3$
and $\mathrm{Var}_{\mathcal{D}_c}[c^2_m] = 4/45$. For off-diagonal
terms, $m \neq m'$, the expectation value is vanishing due to the
symmetry, while the variance scales exponentially as $1/(9\cdot
2^L)$ (computational details are shown in App. \ref{appx: norm}.)
Therefore, using the central limit theorem, we conclude that for
large $M$, the norm $\mathcal{Z}$ is drawn from 
\begin{equation}\label{eq: norm_dist}
   \mathcal{Z} \sim \mathcal{N}\qty(\frac{M}{3}, \frac{4M}{45} 
     + \frac{M(M-1)}{9\cdot2^{L}}).
\end{equation}
In the large-$L$ limit, the off-diagonal terms become negligible,
with the leading behavior governed by the coefficients $c_m$.
Furthermore, we observe that the relative fluctuation of the norm,
$\mathrm{Var}_{\mathcal{D}}[\mathcal{Z}]/\mathbb{E}_{\mathcal{D}}
[\mathcal{Z}]^2$, also vanishes for large $M$. 

\begin{figure}
    \centering
    \includegraphics[width=1.0\linewidth]{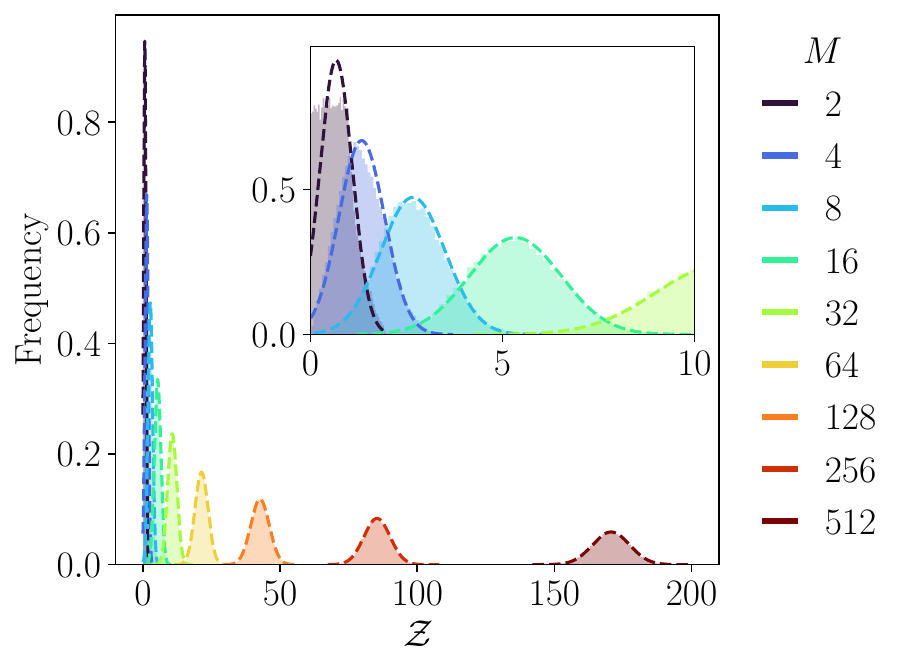}
    \caption{Distributions of 10,000 SPS norms $\mathcal{Z}$ over $\mathcal{D}$ for a system size $L=20$ at different values of $M$. The norms are centered around $M/3$, with the spread increasing as $M$ grows. The inset highlights the small-$M$ regime, where clear deviations from the theoretical Gaussian form can be observed (dashed curves).}
\label{fig: norm}
\end{figure}

Fig.~\ref{fig: norm} demonstrates the normalization factor's
distribution from 10,000 random SPSs. For small $M$, the quantum
state with only a few product states retains an atypical
structure---the norms are correlated sums of a small number of
product-state contributions, and their distribution deviates
substantially from the theoretical Gaussian distributions derived in
Eq.~\eqref{eq: norm_dist} (dashed Gaussian curves in Fig.~\ref{fig:
norm}). When $M$ increases, the distribution converges to the
expected Gaussian distribution.

\subsubsection{Statistics of local observables} 
\label{ssec:statistics_observables}    

\begin{figure}
  \centering
  \includegraphics[width=1.0\linewidth]{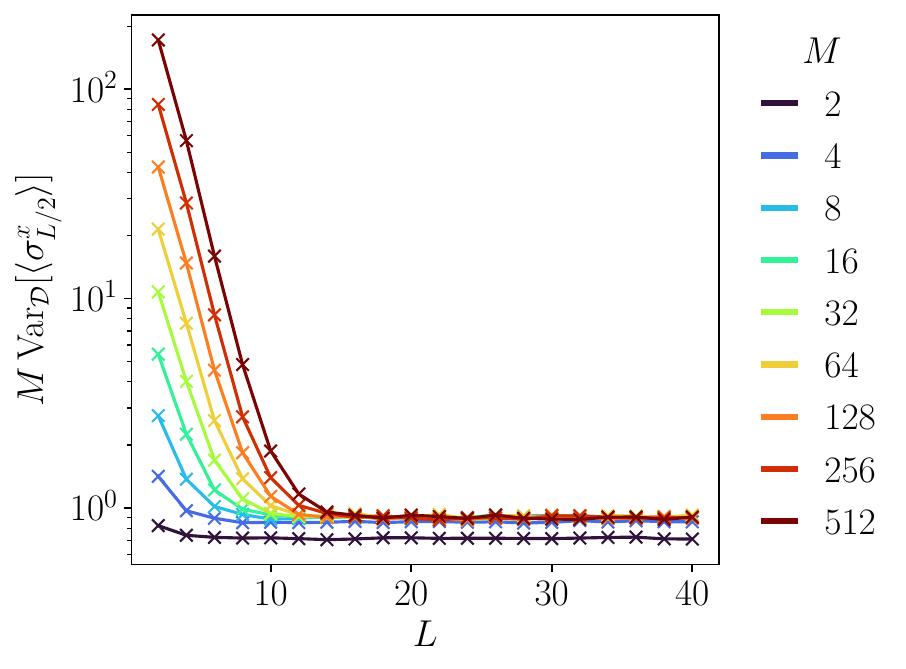}  
  \caption{The variances of 10,000 values of
    $\expval{\sigma^x_{L/2}}$ over $\mathcal{D}$ as a function of
    $L$ for different $M$. The variances remain constant as $L$
    increases. This is an emergence of restricted typicality for the
    SPS ansatz in the large-$M$ limit.}
  \label{fig: paulix}
\end{figure}

We now consider the statistics of the expectation value of a
single-site Pauli-$X$ operator at site $l$, $\sigma^x_l$,
\begin{equation}\label{eq: exp_paulix}
    \expval{\sigma^x_l} = \frac{1}{\mathcal{Z}}\sum_{m,m'=1}^M 
      c_m c_{m'}\sin(\theta_{l}^{m}+\theta_{l}^{m'}) 
      \prod_{k\neq l}^L \cos(\theta_{k}^{m}-\theta_{k}^{m'}),
\end{equation}
derived in App. \ref{appx: pauli}. It is straightforward to verify
that the expectation value $\expval{\sigma^x_l}$ vanishes on
average, since $\mathbb{E}_{\mathcal{D}_\theta^2}
[\mathrm{sin}(\theta^m_l + \theta^{m'}_l)] = 0$ for all $m$ and
$m'$. In Fig.~\ref{fig: paulix}, we show that the variance of
$\expval{\sigma^x_l}$ scales as $1/M$ with increasing $M$, and for
sufficiently large $L$ it becomes effectively independent of the
system size. While the Haar-typical states display a much stronger
concentration, with variances that decay exponentially in $L$
\cite{typicality}, the SPS ansatz displays a complementary and
robust form of \textit{restricted typicality}, where fluctuations
are suppressed polynomially in $M$, not suppressed exponentially.

Furthermore, the independence from $L$ at large system sizes
indicates that the observable retains consistent statistical
behavior even in the thermodynamic limit. This stability also
supports the ansatz’s suitability for variational optimization,
called \textit{trainability}, where fluctuations of gradients do not
vanish exponentially with $L$. We analyze in more detail the
trainability in Sec. \ref{sec: trainability}.

\subsubsection{Typical entanglement entropy} 
\label{ssec:statistics_entanglement}

The evaluation of the entanglement entropy of the SPS ansatz serves
as a key indicator of the \textit{expressive power} of the ansatz.
Entanglement reflects the degree of quantum correlation that can be
represented between subsystems---higher typical entropy indicates
that the ansatz can capture more complex and correlated quantum
states, whereas lower typical entropy suggests a more restricted
expressive power confined to nearly separable configurations. To
analyze the entanglement entropy, we evaluate the 2-R\'enyi entropy,
$S_2(\rho_A) = -\ln\Tr{\rho_A^2}$ of the reduced density matrix
$\rho_A = \Tr_B{\ketbra{\Psi}}$, where subsystem $A$ corresponds to
half of the all system. We focus on the 2-R\'enyi entropy since it
is computationally more accessible than the von Neumann entropy
while retaining the same qualitative behavior. Details for the
computation of the 2-R\'enyi entropy within the framework of the SPS
ansatz are shown in App.~\ref{appx: renyi}. Theoretically, the
maximum of the 2-R\'enyi entropy is reached when the reduced density
matrix becomes maximally mixed over the $M$ product states, yielding
$S_2^\mathrm{max}(\rho_A) = \min\qty{\ln M,\frac{L}{2}\ln2}$ where
$\frac{L}{2}\ln2$ is the maximum entropy for a bipartite system.

\begin{figure}
    \centering
    \includegraphics[width=1.0\linewidth]{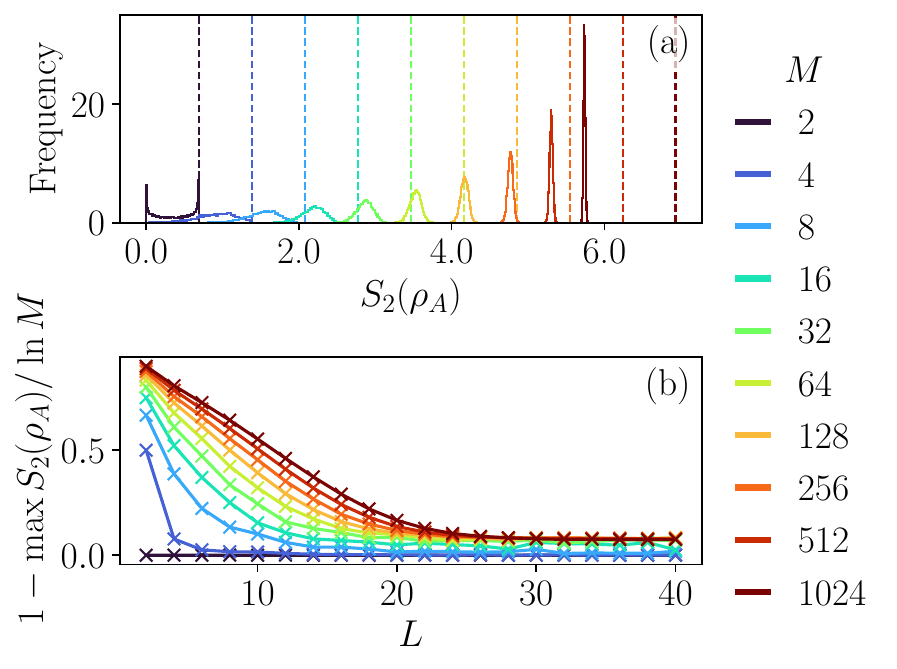}
    \caption{Typical 2-R\'enyi entropies obtained from 10,000 random SPS samples. (a) The distribution of the typical 2-R\'enyi entropies for $L=20$. The entropies typically concentrate far from the theoretical maximum, even as $M$ increases. (b) The minimum relative errors from the theoretical maximum saturate for larger $L$.}
    \label{fig: ent_ent}
\end{figure}

Fig.~\ref{fig: ent_ent}(a) demonstrates that the typical 2-R\'enyi
entropies, obtained from 10,000 random SPS realizations, become
increasingly concentrated as $M$ grows.  However, Fig.~\ref{fig:
ent_ent}(a) also reveals that the typical entropies remain below the
theoretical maximum given by $\ln M$ and depicted by the dashed
lines. Fig.~\ref{fig: ent_ent}(b) illustrates that the relative
distance between the typical entropy and the maximum entropy does
not vanish, but rather saturates for large $L$, even as $M$
increases. In contrast to the SPS behaviour, a random MPS with local
dimension equal to 2 and bond dimension $\chi$ obeys an area law
with high probability. Ref. \cite{collins2013matrix} and Theorem 4.7
of Ref. \cite{haferkamp2021emergent} show that the 2-R\'enyi entropy
for a single bipartition of an open chain, the typical half-chain
entropy satisfy $S_{2}\approx \min\qty{\ln\chi,\frac{L}{2}\ln
2}+O(\chi^{-1})$. This distinction highlights that while random MPS
can reach the entanglement capacity allowed by their bond dimension,
the SPS ansatz is inherently more restricted, which directly affects
their expressive power and typical entanglement structure. This
comparison confirms that the numerical analysis characterizes this
limitation by revealing how the typical entanglement saturates below
the theoretical bound.

\begin{figure*}
    \centering
    \includegraphics[width=1.0\linewidth]{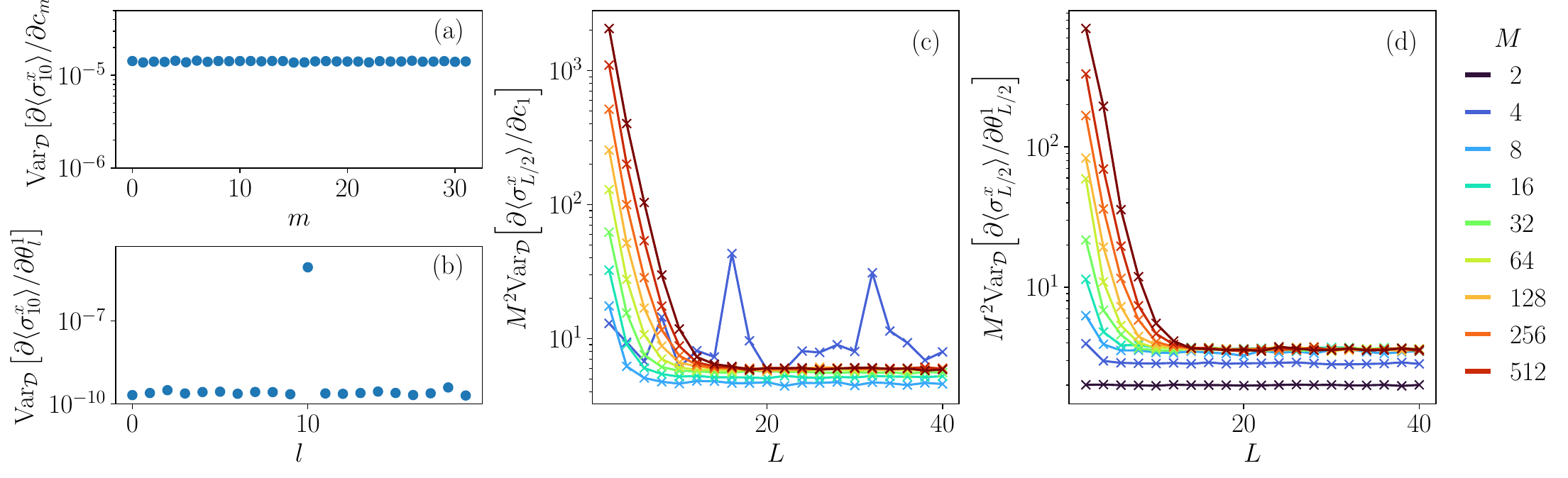}
    \caption{Trainability of the SPS ansatz is analyzed by evaluating the gradients of a local observable with respect to the variational parameters across 10,000 random samples. Panels (a) and (b) correspond to the case of $L=20$ and $M=32$. In panel (a), the gradients with respect to the coefficients $c_m$ are identical across all $m$. In panel (b), fixing $m=1$, the gradients with respect to the site index $l$ vanish everywhere except at the site where the local observable is applied. Panels (c) and (d) display the gradient variances as functions of system size $L$ and the number of product states $M$, focusing on derivatives with respect to $c_1$ and $\theta^1_{L/2}$, respectively. Although the case $M=2$ in panel (c) shows fluctuations due to finite-sample effects, overall the variances decrease and saturate for large $L$, while decaying polynomially with $M^2$. These results demonstrate that the SPS ansatz avoids barren plateaus, thereby ensuring efficient optimization even for large quantum systems.
}
    \label{fig: grad}
\end{figure*}

\subsection{Trainability}\label{sec: trainability} 
\label{ssec:trainability}

A main challenge in variational quantum algorithms is the \textit{barren
plateau} problem~\cite{mcclean2018barren,cerezo2021variational,larocca2025barren,srimahajariyapong2025connecting}, where the gradient variance of loss functions
decays exponentially with the system size~\cite{barthel2025absence},
making optimization prohibitively difficult.  This behavior
typically emerges in highly entangled or randomly initialized
ans\"atze, where the concentration of measure suppresses local
gradient signals throughout the Hilbert space. 

Since we are interested in searching for ground states, with a loss
function corresponding to the energy of the system, we here examine
the variance of the gradients of a local observable. Consider the
ansatz in an $L$-site system; it is practical to focus on
representative derivatives. We begin with the derivative of the
site-$L/2$ expectation value $\langle\sigma^x_{L/2}\rangle$ with
respect to the coefficients $c_m$, shown in Fig.~\ref{fig: grad}(a)
for $L=20$ and $M=32$ with $10,000$ different realizations. 

The
variances of the gradients $\partial
\langle\sigma^x_{10}\rangle/\partial c_m$ have not vanished across
different $m$, indicating that these gradients remain essential
across terms, thus shifting our focus towards $\theta^m_l$. 
Next, we evaluate the derivatives with respect to $\theta^1_l$
across all sites while considering only a term $m=1$. As shown in
Fig.~\ref{fig: grad}(b), all gradients $\partial
\langle\sigma^x_{10}\rangle/\partial \theta^1_l$ vanish except at
the site directly coupled to the observable $\sigma^x_{10}$. This
behavior indicates that although many subspaces appear flat and
could resemble barren plateaus, the relevant local subspace
associated with the observable retains finite gradients, thereby
ensuring that optimization remains feasible in practice. 
We remark that, for the random nature of the SPS analyzed, the same results occur also when one considers local observables at other sites.  

To gain more insight, we examine the variances of
$\partial\langle\sigma^x_{L/2}\rangle/\partial c_1$ and
$\partial\langle\sigma^x_{L/2}\rangle/\partial \theta^1_{L/2}$ as
functions of the system size $L$ for different $M$. Figs.~\ref{fig:
grad}(c,d) reveal that both variances scale as $1/M^2$ with
increasing $M$, and eventually collapse on a single curve as $L$
grows larger, even when $M$ continues to increase. This scaling
behavior highlights that the SPS ansatz retains trainability at
large system sizes, in sharp contrast to the exponential decay
characteristic of barren plateaus. This locality property further
implies that the derivatives with respect to all $\theta^m_l$ remain
non-vanishing for Hamiltonians composed of local observables. Hence,
this observation provides strong evidence for the \textit{absence}
of a barren plateau in the SPS ansatz, making the ansatz
\textit{trainable}. Although this analysis does not directly assess
task-specific optimization, it only provides strong evidence that
the SPS ansatz possesses favorable gradient landscapes that support
efficient training in practice.

\section{Ground state search by SPS ansatz}
\label{sec: GS}

It has been established that the SPS ansatz is \textit{trainable},
as its optimization does not suffer from barren plateaus. In this
section, we present numerical results for the ground-state energies
of tilted Ising models on various lattice geometries using the SPS
ansatz. To identify the ground state within the SPS framework, we
minimize the variational energy, 
\begin{equation}
    E_{\text{SPS}}(\Theta) = \mel{\Psi}{\mathcal{H}}{\Psi},
\end{equation}  
by optimizing all variational parameters $\Theta = \{c_m,
\theta_l^{m}\}$ where initially drawn from the uniform distributions in Sec.~\ref{ssec:statistics_wave}. Due to the product-state structure of the ansatz,
the loss function can be expressed analytically as a function of
trigonometric combinations of $\Theta$ (see App. \ref{appx:
pauli}.) This analytic tractability can significantly speed up the
optimization compared to fully black-box approaches (see Sec.
\ref{sec: compare}.)

\begin{figure}
    \centering
    \includegraphics[width=1.0\linewidth]{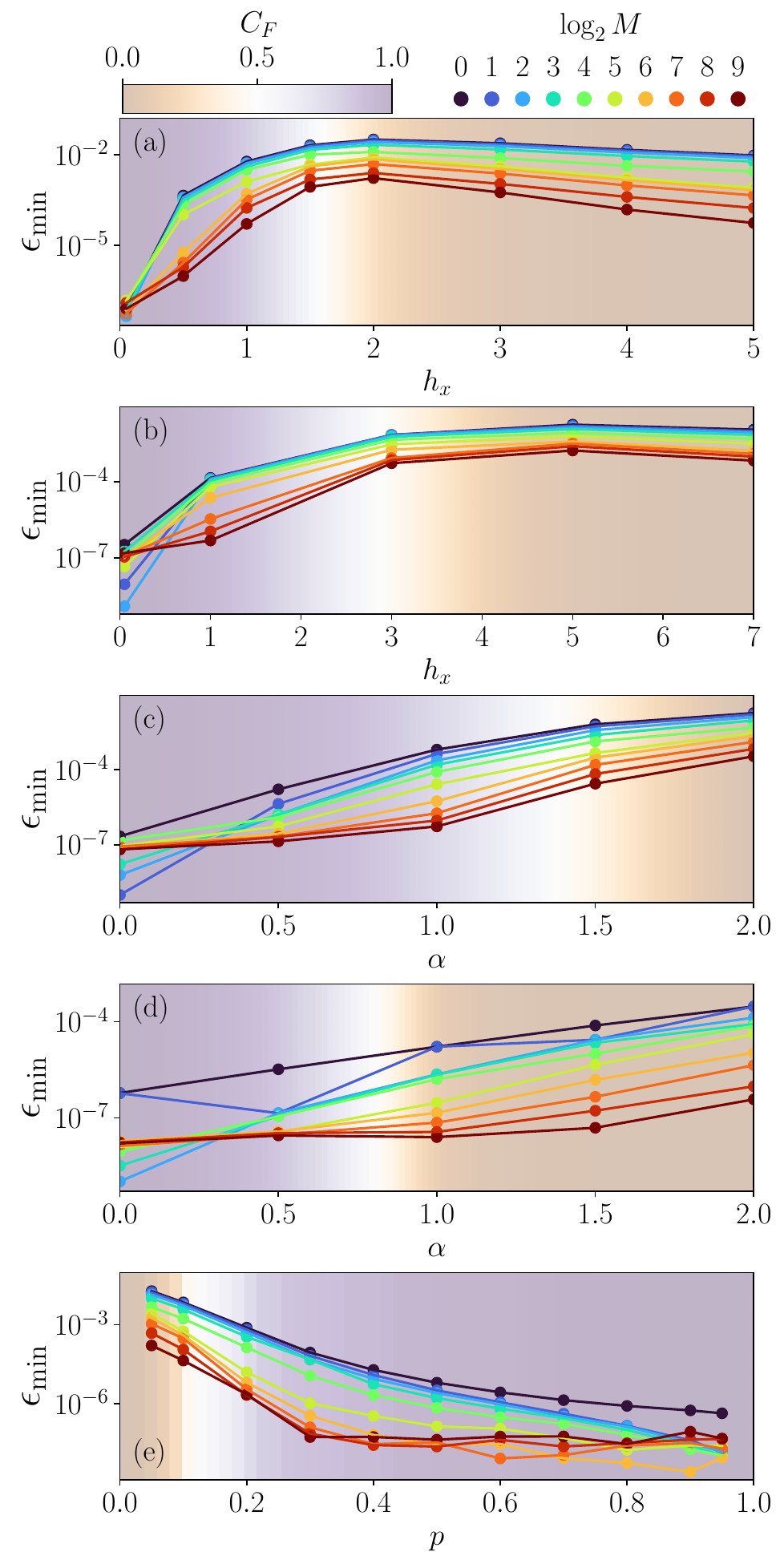}
    \caption{The performances in the ground state searches task with the SPS ansatz. The performances with different $M\in \{1,2,4,\ldots,512\}$ are quantified by the minimum relative error, from 20 different initialized SPS parameters, in the SPS ansatz’s approximation of the ground state energy compared to the target ground state energy $E_{\text{GS}}$ for a tilted Ising model with $J=-1$, $h_z = 0.25$ on (a) 40-qubit 1D nearest-neighbor connection for $h_x \in \{0.05,0.5,1.0,1.5,2.0,3.0,4.0,5.0\}$, (b) ($4\times4\times4$)-qubit 3D nearest-neighbor connection for $h_x = \{0.05,1.0,3.0,5.0,7.0\}$, (c) 40-qubit 1D all-to-all connection for $\alpha\in\{0.0,0.5,1.0,1.5,2.0\}$ with $(J,h_x,h_z) = (-1,3.0,0.25)$, (d) ($4\times4\times4$)-qubit 3D all-to-all connection for $\alpha\in\{0.0,0.5,1.0,1.5,2.0\}$ with $(J,h_x,h_z) = (-1,7.0,0.25)$, and (e) 40-qubit random connected graph for $p\in\{0.05,0.1,0.2,0.3,0.4,0.5,0.6,0.7,0.8,0.9,0.95\}$. The gradient background is shaded according to the ferromagnetic correlator $C_F$, with lighter shading corresponding to lower values.} 
    \label{fig: GS}
\end{figure}

We focus on finding the ground-state energy of the tilted Ising
model in the presence of a transverse field $h_x$ and longitudinal
field $h_z$, given by
\begin{equation}
  \mathcal{H} = \sum_{\expval{k, l}}^{L} J_{kl}
    \sigma^z_{k}\sigma^z_{l} - h_x\sum_{k=1}^L \sigma^x_{k} 
      - h_z\sum_{k=1}^L\sigma^z_{k},
\end{equation}
where $\expval{k, l}$ denotes all connected pairs in an arbitrary
lattice with $L$ spins with coupling $J_{kl}$. We will consider
nearest neighbour coupling in 1D and 3D configurations, 1D and 3D
systems with power-law long-range interactions and a system with
all-to-all coupling but such that only a random set of sites is
coupled. To evaluate the quality of the ground state, we will plot the
relative error from an accurate computation done with matrix product
states---we will use the DMRG algorithm with a snake configuration
of an MPS. To gain deeper insights into the performance of the SPS
ansatz, we evaluate the ferromagnetic correlator 
\begin{equation}\label{eq: C_F}
  C_F = \frac{1}{L-1} \sum_{l \neq L/2}^{L} \expval{\sigma^z_{L/2} 
    \sigma^z_l}-\expval{\sigma^z_{L/2}}\expval{\sigma^z_l}. 
\end{equation}  
In this study, we focus exclusively on transitions between
paramagnetic and ferromagnetic phases, by considering $J < 0$. This
correlator has been computed using a DMRG ground state search
algorithm. A ferromagnetic ground state is characterized by a large
value of the ferromagnetic correlator $C_F$ (indicated by the shaded
purple region in Fig.~\ref{fig: GS}), whereas a paramagnetic ground
state driven by the transverse field yields a small correlator value
(shaded orange). The portions of Fig.~\ref{fig: GS} with a white
background are the region close to the phase transition boundary. 

\subsection{Optimization} 

In our implementation, we employ the AdamW optimizer \cite{AdamW}
with analytic gradients of the loss function $\partial
E_\mathrm{SPS}(\Theta)/\partial \Theta$ (see App. \ref{app: grad})
to iteratively refine the SPS parameters until convergence is
achieved. We note, however, that when examining the coefficients
$\{c_m\}$ in the paramagnetic regime, at times we might observe that
several of them become negligibly small, which then results in the
optimization being trapped in a local minimum. We counter this by
implementing a \textit{resampling} strategy. Specifically, in
certain optimization steps, whenever any coefficients $c_m$ fall
below a predetermined threshold, we reinitialize both $c_m$ and its
corresponding set of angles $\{\theta^{m}_l\}_{l=1}^L$, drawn uniformly from the same distributions in Sec. \ref{ssec:statistics_wave}. This approach helps to avoid wasting computational resources on
product states that contribute little to the overall wavefunction,
and it provides an opportunity to explore more promising regions of
the parameter space.

In practice, the learning rate of the AdamW algorithm is $10^{-3}$. The optimization epochs for each sample are 20,000 steps, and the resampling on the states with $c_m<10^{-5}$ occurs every 5,000 epochs; thus, the resampling criteria happen 4 times in the optimization.

\subsection{Implementation}

In our implementation, the SPS ansatz is constructed and optimized
within the PyTorch framework, which naturally supports GPU
acceleration and parallelization, enabling efficient simulations of
large system sizes. All SPS simulations are executed on nodes
equipped with NVIDIA A100 Tensor Core GPUs \cite{NSCC}. The
benchmark ground-state energies are obtained using the DMRG
algorithm, implemented in the TeNPy package
\cite{tenpy2024}, which is executed on CPUs
AMD~EPYC$^{\mathrm{TM}}$~7713 processors \cite{NSCC}, where even for
higher-dimensional lattices, the geometry is mapped into a
one-dimensional snake-like configuration of MPS.

From the perspective of computational complexity, a single DMRG
sweep over the chain scales as $O(L\chi^3)$ in time and $O(L\chi^2)$
in memory, where $\chi$ is the bond dimension of the MPS. Since
practical convergence requires $N_\text{sweep}$ full sweeps, the
overall scaling is $O(N_\text{sweep}L\chi^3)$ in time. This makes
DMRG highly efficient for systems with area-law entanglement but
increasingly expensive for critical or higher-dimensional systems.
By contrast, the SPS ansatz has a memory scaling as $O(LM)$,
computes the expectation value and gradients of the Hamiltonian in
$O(LM^2)$, and optimizes the SPS parameters in $O(LM^2T)$ where $T$
is the number of optimization steps. The calculation details of the complexity
are shown in Apps. \ref{appx: pauli} and \ref{app: grad}.

\begin{table*}[]
\centering

\caption{Comparison between DMRG and SPS methods in terms of computational times for different systems. For DMRG, the bond dimension with the lowest energy $E_\mathrm{GS}$ is denoted by $\chi*$ and $\chi$ for the bond dimension that the energy has a relative error $\epsilon_\chi$ compared with $E_\mathrm{GS}$. While for SPS, $M$ represents the number of product states and $\epsilon$ the relative error compared with $E_\mathrm{GS}$. The listed systems include 40-qubit one-dimensional (1D) and $(4\times 4 \times 4)$-qubit three-dimensional (3D) nearest-neighbor (NN) models, 40-qubit 1D and $(4\times 4 \times 4)$-qubit 3D long-range (LR) models with a specific $\alpha = 2.0$, and a 40-qubit random-coupling (RC) model with a specific $p = 0.4$. In this comparison, the system parameters are $(J,h_x,h_z) = (-1.0,3.0,0.25)$ for 1D and RC systems and $(-1.0,7.0,0.25)$ for 3D systems. }
\begin{tabular}{|c||ccccc||cccc|}
\hline
\multirow{2}{*}{System} & \multicolumn{5}{c||}{DMRG}                                                                                                                                   & \multicolumn{4}{c|}{SPS}                                                                                                       \\ \cline{2-10} 
                        & \multicolumn{1}{c|}{$\chi*$} & \multicolumn{1}{c||}{$E_\mathrm{GS}$}     & \multicolumn{1}{c|}{$\chi$} & \multicolumn{1}{c|}{$\epsilon_{\chi}$}   & Time (s) & \multicolumn{1}{c|}{$M$} & \multicolumn{1}{c|}{$\epsilon$}    & \multicolumn{1}{c|}{CPU time (s)}     & GPU time (s)    \\ \hline
1D NN                   & \multicolumn{1}{c|}{11}      & \multicolumn{1}{c||}{-3.10379965(1)} & \multicolumn{1}{c|}{2}      & \multicolumn{1}{c|}{$1.07\times10^{-6}$} & 8.21     & \multicolumn{1}{c|}{512} & \multicolumn{1}{c|}{$1.00\times 10^{-3}$} & \multicolumn{1}{c|}{$278.0 \pm 9.6$}            & $7.08 \pm 0.25$ \\ \hline
3D NN                   & \multicolumn{1}{c|}{512}     & \multicolumn{1}{c||}{-7.09401002(1)}   & \multicolumn{1}{c|}{8}      & \multicolumn{1}{c|}{$1.89\times10^{-3}$} & 72.77    & \multicolumn{1}{c|}{100} & \multicolumn{1}{c|}{$1.89\times10^{-3}$}  & \multicolumn{1}{c|}{$27.47 \pm 9.36$} & $1.81 \pm 0.44$ \\ \hline
1D LR                   & \multicolumn{1}{c|}{105}     & \multicolumn{1}{c||}{-3.16470959(4)} & \multicolumn{1}{c|}{2}      & \multicolumn{1}{c|}{$2.86\times10^{-4}$} & 13.06    & \multicolumn{1}{c|}{256} & \multicolumn{1}{c|}{$1.00\times 10^{-3}$} & \multicolumn{1}{c|}{$471.4 \pm 139.0$}            & $7.37 \pm 0.37$ \\ \hline
3D LR                   & \multicolumn{1}{c|}{128}     & \multicolumn{1}{c||}{-9.53603316(8)}  & \multicolumn{1}{c|}{4}      & \multicolumn{1}{c|}{$3.71\times10^{-5}$} & 243.3    & \multicolumn{1}{c|}{128} & \multicolumn{1}{c|}{$3.71\times10^{-5}$}  & \multicolumn{1}{c|}{$140.7 \pm 84.6$}            & $3.87 \pm 1.77$ \\ \hline
RC                      & \multicolumn{1}{c|}{128}     & \multicolumn{1}{c||}{-8.65942219(1)}  & \multicolumn{1}{c|}{2}      & \multicolumn{1}{c|}{$1.16\times10^{-5}$} & 17.24    & \multicolumn{1}{c|}{32}  & \multicolumn{1}{c|}{$1.16\times10^{-5}$}  & \multicolumn{1}{c|}{$1.05 \pm 0.74$}  & $0.65 \pm 0.41$ \\ \hline
\end{tabular}
\label{tab: compare_table}
\end{table*}

\subsection{Nearest-neighbor interactions} 

For 1D and 3D lattices with nearest-neighbor interactions and having
open boundary conditions, we consider uniform coupling strengths
defined by $J_{ij} = J$. In our experiments, we optimize the energy
$E_{\text{SPS}}$ with different $M=1,2,4,\ldots,512$ for two models:
a 40-qubits 1D ferromagnetic tilted Ising model with $J < 0 $ and
$h_z = 0.25|J|$ while varying $h_x/|J|$, and a $(4 \times 4 \times
4)$-qubit 3D tilted Ising model on a cube lattice. 

In practice, we first benchmark the SPS ansatz on a 1D system by
initializing 20 random instances and evaluating their performance.
As shown in Fig.~\ref{fig: GS}(a), the SPS ansatz can approximate
the ground-state energy in the ferromagnetic phase with remarkable
accuracy. The relative error, defined as 
\begin{equation}\label{eq: rel_err}
    \epsilon :=\frac{{E_{\text{SPS}} - E_{\text{GS}}}}{E_{\text{GS}}},
\end{equation}
is small also when compared against the ground-state energy
$E_{\text{GS}}$ obtained from the DMRG algorithm. For small $h_x$, a single product state ($M=1$), the minimum relative error
$\epsilon_{\min}$ can reach values as low as $10^{-8}$. As the
transverse field $h_x$ increases, however, the accuracy depends more
strongly on the number of product states $M$. Near the critical
region, the performance with small $M$ is significantly reduced,
consistent with the required growth of the bond dimension
in a DMRG algorithm at criticality. 

In the paramagnetic phase, optimization with larger $M$ yields
moderate improvements, and while the achieved accuracy is lower than
in the ferromagnetic regime, the relative errors remain fairly
small, underscoring the robustness of the ansatz.
Ref.~\cite{florido2024product} showed that translationally invariant
MPSs can always be written either as a single product state or as a
superposition of only a few such states. Since the paramagnetic
ground state is effectively translationally invariant only in the
bulk, the observed performance degradation can be attributed
primarily to boundary effects.

We further test the SPS ansatz on a 3D system. The results, shown in
Fig.~\ref{fig: GS}(b), demonstrate that the SPS ansatz maintains
good performance in the low-$h_x$ regime, but its accuracy
deteriorates more noticeably as the transverse field strength
increases, particularly in the vicinity of the phase transition
($h_x \approx 2.8$). This behavior highlights both the expressive
power and the limitations of the SPS ansatz---it efficiently
captures ground states in the ferromagnetic regions, but requires larger $M$ to
remain accurate in the critical region or for the paramagnetic regions.

\subsection{Long-range and random interactions} 

To further investigate the performance of the SPS ansatz in
ground-state searches, we carry out additional numerical experiments
on tilted Ising models with long-range and randomly coupled
interactions. For the long-range interaction, we will consider an
experimentally relevant one (e.g. in trapped ions) that can be
modeled using a power-law decay: $J_{kl} = J/d(k,l)^\alpha$, where
$J$ sets the interaction strength, $d(k,l)$ is an Euclidean distance
between site $k$ and $l$ on a lattice where each nearest-neighbor
pair has an unit length proximity, and $\alpha$ determines the
spatial decay of correlations. For the random coupling system, a
pair of sites is connected with probability $p$ and interacted with
uniform strength $J$. The model parameters are fixed to $J<0$,
$h_x=3.0|J|$, and $h_z=0.25|J|$ for 1D and random coupling systems,
and $h_x=7.0|J|$ with the same $J$ and $h_z$ for a 3D system,
corresponding to the paramagnetic regime in the short-range limit
(large $\alpha$ in long-range system). In these experiments, we
systematically vary both the interaction-range exponent $\alpha$ and
the coupling probability $p$, respectively. Note that we use DMRG
algorithms to obtain the ground-state energies in practice and use
the snake configuration for a 3D system.

Figs.~\ref{fig: GS}(c,d) show the minimum relative error
$\epsilon_\mathrm{min}$ for each interaction range exponent
$\alpha$, evaluated over 20 different random initializations on
40-qubit 1D and $(4\times 4\times 4)$-qubit 3D long-range systems,
respectively. When $\alpha = 0$, corresponding to a fully connected
interaction, the SPS ansatz achieves low relative errors, with
$\epsilon_{\min}$ reaching the order of $10^{-8}$. As $\alpha$
increases---resulting in shorter-range interactions---the minimum
relative error $\epsilon_{\min}$ also becomes larger. This trend
mirrors changes in the system's ferromagnetic correlator
$C_F$~\eqref{eq: C_F}, which is illustrated as the background color
in Figs.~\ref{fig: GS}(c,d). Similar to the nearest-neighbor case,
we observe that lower ferromagnetic correlations are associated with
higher relative errors. 
 
For the 40-qubit tilted Ising model defined on a random-connecting
graph, where interactions are disordered and each edge is present
with probabilities $p \in
\{0.05,0.1,0.2,0.3,0.4,0.5,0.6,0.7,0.8,0.9,0.95\}$, the SPS ansatz
continues to perform effectively. In practice, we use the same
random seed in every random graph generation, and the are adjacency matrices are given in App.~\ref{app: conn}. As shown in
Fig.~\ref{fig: GS}(e), even in this highly irregular interaction
landscape, relative errors as small as $10^{-5}$ are achieved with
around $M \sim 100$ product states. This result highlights the
robustness of the SPS framework---despite the absence of
translational symmetry and the presence of disorder, the ansatz can
still capture the essential features of the ground state with
moderate resources. In particular, the ability to handle disordered
systems suggests that the SPS ansatz may be a flexible variational
tool beyond clean, uniform models. 

For $\alpha = 0$ in long-range systems, and large $p$, e.g., $p \geq 0.6$,
in random coupling, the systems become (almost) uniformly
all-to-all interaction, yielding a ground state close to a
low-entanglement state. Accordingly, a small $M$ in the SPS ansatz
already captures the essential structure, while a larger $M$
primarily increases the parameter space and hinders optimization
under a fixed compute budget. As a result, within the same training
time, large $M$ can exhibit higher relative errors than small $M$,
reflecting the optimization effect rather than limited expressivity.

\subsection{Comparison}\label{sec: compare}

To have a quantitative benchmark of the SPS ansatz, we compare its ability
to reach the ground-state energy with that of the DMRG algorithm
under the same relative-error target (when obtainable within a manageable value of $M$, i.e. beyond 1D). We focus on the paramagnetic regime for which the SPS ansatz generally requires larger values of $M$, compared to small values for $\chi$ in the DMRG code. The ground-truth energies,
$E_\mathrm{GS}$, are obtained from DMRG at a bond dimension $\chi^*$
chosen such that the energy change between consecutive doublings of the bond dimension ($\chi = 2,4,8,\ldots$) is below $10^{-10}$; see Table~\ref{tab: compare_table}.

We consider system sizes matching those in the previous section and
use the following parameters: $(J,h_x,h_z)=(-1,\,3.0,\,0.25)$ for 1D
and random-coupling systems, and $(J,h_x,h_z)=(-1,\,7.0,\,0.25)$ for
3D systems. For long-range interactions we set $\alpha=2.0$, and for
random-coupling systems we use $p=0.4$. The SPS ansatz is optimized
with AdamW (learning rate $\eta=0.1$) and training stops once the
SPS energy $E_{\mathrm{SPS}}$ reaches the prescribed relative-error
threshold. The SPS runtimes reported in Table~\ref{tab:
compare_table} are means $\pm$ standard deviations over 40
realizations where the initial SPS parameters are drawn from the distribution described in Sec. \ref{ssec:statistics_wave}.

To ensure a fair comparison, we benchmark DMRG and SPS on identical
hardware: first on a CPU-only node with an AMD EPYC$^{\text{TM}}$
7713 processor. We additionally show the computational time on a GPU
node equipped with an NVIDIA A100~\cite{NSCC}. In 3D and
random-coupling systems, the SPS ansatz typically reaches the target
relative error faster than DMRG, with further speedups observed on
the GPU. In contrast, for 1D systems DMRG remains highly efficient
with short runtimes, whereas the SPS ansatz often does not attain
the same relative-error threshold, reflecting the optimality of DMRG for 1D systems.

\section{Conclusions}\label{sec: conclusion}

The SPS ansatz provides a compact variational framework built from a
superposition of product states, directly linked to a rank-$M$ CP
decomposition of a quantum state tensor, making the ansatz
efficient, scalable, and highly parallelizable.  We found that the
ansatz exhibits restricted typicality, where local observables are
partially stable across random instances, but their variances
display polynomial decreases on $M$ instead of exponential scaling.
Regarding the expressive power, we have shown that for a random SPS,
the 2-R\'enyi entropy of the reduced density matrix exhibits an
amount of entanglement comparable to the maximum expressible by the
ansatz. A central strength of the SPS ansatz is its trainability,
whereby analysis of the variance of the grandients confirm the
absence of barren plateaus.  Local observables retain non-vanishing
gradients, and their variances decay polynomially with $M$ and
saturate with $L$, guaranteeing efficient optimization even in large
systems.  Despite its simplicity, we have shown that the SPS ansatz
is expressive enough to approximate nontrivial quantum correlations
and achieves strong performance in ground-state searches. It
captures ordered and paramagnetic phases with high accuracy, and
remains effective in disordered settings such as long-range and
random-coupling tilted Ising models.

The limited performance of the SPS ansatz in the paramagnetic phase
suggests that incorporating spatial modulation in the ansatz could
be a natural and promising extension. By allowing site-dependent
structure in the variational parameters, the ansatz can capture weak
correlations beyond mean-field descriptions, thereby broadening its
expressive power in more strongly correlated regimes. Future works
would also focus on extending the ansatz to complex-valued quantum
states, for the evaluation of more complex ground states, and for
the description of time-evolving systems.

\section*{Acknowledgment} 
D.P. and A.S. acknowledges the support of the Ministry of Education,
Singapore, under the grant T2EP50123-0017, and from HTX under
project HTX000ECI24000267. D. P. acknowledges fruitful discussions
with D. Rossini and G. Vignale.  The computational work was
performed at the National Supercomputing Centre, Singapore
\cite{NSCC}.

\bibliography{ref}

\appendix 

\section{Analytical formulations}\label{appx: analytic} 

A notable advantage of the SPS ansatz is its analytic tractability.
In contrast to many variational families that require heavy
numerical sampling, several physical quantities can be computed
explicitly. Now, we present the analytic formulas for the mean and
the variance of the norm $\mathcal{Z}$, for the 2-R\'enyi entropy, for local observables within the SPS framework, and for the gradients of the local observables with respect to the SPS parameters.

\subsection{Mean and variance of an SPS norm}\label{appx: norm}

Starting from the SPS norm $\mathcal{Z}$, each parameter is independent, allowing us to compute the expectation values of individual terms separately.
For the diagonal term, since the probability density function of $c_m$ is uniform with $p(c_m) = 1/2$, the expectation value of $c_m^2$ is
\begin{equation}
    \mathbb{E}_{\mathcal{D}_c}[c_m^2]
    = \frac{1}{2}\int_{-1}^{1} c_m^2\, \mathrm{d}c_m
    = \frac{1}{3}.
\end{equation}
Similarly, the variance of $c_m^2$ is
\begin{align}
    \mathrm{Var}_{\mathcal{D}_c}[c_m^2]
    &= \frac{1}{2}\int_{-1}^{1} c_m^4\, \mathrm{d}c_m
       - \bigl(\mathbb{E}_{\mathcal{D}_c}[c_m^2]\bigr)^2 \nonumber\\
    &= \frac{4}{45}.
\end{align}
Since there are $M$ diagonal terms, their total variance contribution is $4M/45$.

For the off-diagonal terms, the expectation value vanishes due to the symmetry of the cosine function.
Given that the probability density of each $\theta^m_l$ is $p(\theta^m_l) = 1/\pi$, we have
\begin{align}
    \mathrm{Var}_{\mathcal{D}_\theta^2}&\qty[\cos\qty(\theta^m_l - \theta^{m'}_l)] \nonumber\\
    &= \frac{1}{\pi^2}
       \int_{-\frac{\pi}{2}}^{\frac{\pi}{2}}\!
       \int_{-\frac{\pi}{2}}^{\frac{\pi}{2}}
       \cos^2\qty(\theta^m_l - \theta^{m'}_l)
       \, \mathrm{d}\theta^m_l \, \mathrm{d}\theta^{m'}_l \nonumber\\
    &= \frac{1}{2},
\end{align}
and
$\mathrm{Var}_{\mathcal{D}_c^2}[c_m c_{m'}] = 1/9.$
Since each off-diagonal term of $\mathcal{Z}$ involves $L$ multiplicative factors and $M(M-1)$ summations,
the total variance contribution becomes $M(M-1)/(9 \cdot 2^L)$.
Thus, the overall variance of the SPS norm can be written as
\begin{equation}
    \mathrm{Var}_\mathcal{D}[\mathcal{Z}]
    = \frac{4M}{45} + \frac{M(M-1)}{9 \cdot 2^L}.
\end{equation}

\subsection{2-R\'enyi entropy}\label{appx: renyi}
Considering the reduced density matrix $\rho_A$ of a bipartition of
the system, within the SPS ansatz framework, we can simply write the
reduced density matrix as
\begin{equation}
    \rho_A = \frac{1}{\mathcal{Z}}\sum_{m,m'=1}^M c_mc_{m'}\bigotimes_{l=1}^{L/2}\ketbra{\theta^m_l}{\theta^{m'}_l} C^{mm'}_{B},
\end{equation}
where $C^{mm'}_{B} = \displaystyle\prod_{l=L/2+1}^{L} \braket{\theta^m_l}{\theta^{m'}_l}$ and $\braket{\theta^m_l}{\theta^{m'}_l} = \cos(\theta^m_l - \theta^{m'}_l)$. Then the trace of the square of the reduced density matrix can also be written in the same way,
\begin{equation}
    \Tr\rho_A^2 = \frac{1}{\mathcal{Z}^2}\sum_{m,m',n,n'=1}^M c_mc_{m'}c_nc_{n'}C^{mm'}_{B}C^{m'n}_{A}C^{nn'}_{B}C^{n'm}_{A}
\end{equation}
where $C^{mm'}_{A} = \displaystyle\prod_{l=1}^{L/2} \braket{\theta^m_l}{\theta^{m'}_l}$. In practice, we can implement this quantity using 
\begin{equation}\label{eq: tr_rho^2}
    \Tr\rho_A^2 = \frac{1}{\mathcal{Z}^2}\Tr(\mathcal{C}_B\mathcal{C}_A\mathcal{C}_B\mathcal{C}_A)
\end{equation}
where $\mathcal{C}_A = (C^{mn}_A)_{mn}$ and $\mathcal{C}_B =
(c_mc_nC^{mn}_B)_{mn}$.  From Eq.~\eqref{eq: tr_rho^2}, we can thus
readily evaluate the 2-R\'enyi entropy, which is simply written as
$S_2(\rho_A) = -\ln\Tr{\rho_A^2}$.

In the complexity analysis, building each contraction matrix 
$\mathcal{C}$ requires $O(LM^2)$ operations and $O(M^2)$ memory. The
final evaluation of Eq.~\eqref{eq: tr_rho^2} then scales as $O(M^2
(L+ M))$ with the same memory usage, where the $O(M^3)$ term
originates from the dense matrix multiplications.

\subsection{Local Observables}\label{appx: pauli}

In the analyses of typicality, trainability, and ground-state search
with the SPS ansatz, the computation relies on expectation values of
local observables. This section introduces the analytic formulations
of these local observables.  

Considering, for instance, the Pauli-$X$ operator at site
$k$, $\sigma^x_k$, the expectation value is given by
\[
\expval{\sigma^x_k} = \frac{1}{\mathcal{Z}}\mel{\Psi_u}{\sigma^x_k}{\Psi_u}.
\]  
At the site $k$, the matrix element takes the form 
\begin{align*}
    \mel{\theta^m_k}{\sigma^x_k}{\theta^{m'}_k} 
    &= \begin{pmatrix}\cos\theta^m_k & \sin\theta^m_k\end{pmatrix}
       \begin{pmatrix}0 & 1 \\ 1 & 0\end{pmatrix}
       \begin{pmatrix}\cos\theta^{m'}_k \\ \sin\theta^{m'}_k\end{pmatrix} \\
    &= \cos\theta^m_k \sin\theta^{m'}_k + \sin\theta^m_k \cos\theta^{m'}_k \\
    &= \sin(\theta^m_k + \theta^{m'}_k)
\end{align*}  
and otherwise are the overlap terms $\langle \theta^m_l |
\theta^{m'}_l \rangle = \mathrm{cos}(\theta^m_l - \theta^{m'}_l)$. 
Defining $C^{mm'} = \prod_{l=1}^L \langle \theta^m_l | \theta^{m'}_l
\rangle$, the expectation value can be written compactly as 
\begin{equation}\label{eq: pauli-X}
    \expval{\sigma^x_k} = 
    \frac{\displaystyle \sum_{m,m'=1}^M c_m c_{m'}\, C^{mm'} X^{mm'}_k}
         {\displaystyle \sum_{m,m'=1}^M c_m c_{m'}\, C^{mm'}}
\end{equation}  
where  
\[
X^{mm'}_k = \frac{\sin(\theta^m_k + \theta^{m'}_k)}{\displaystyle\cos(\theta^m_k - \theta^{m'}_k)},
\] 
called the \textit{Pauli-$X$ tensor}, which is the same form shown in Eq.~\eqref{eq: exp_paulix}. This
formulation enables us to compute local expectation values with a
time complexity of $O(LM^2)$.  

For $\sigma^z_k$ operator, the corresponding local term is  
\begin{align*}
    \mel{\theta^m_k}{\sigma^z_k}{\theta^{m'}_k} 
    &= \begin{pmatrix}\cos\theta^m_k & \sin\theta^m_k\end{pmatrix}
       \begin{pmatrix}1 & 0 \\ 0 & -1\end{pmatrix}
       \begin{pmatrix}\cos\theta^{m'}_k \\ \sin\theta^{m'}_k\end{pmatrix} \\
    &= \cos\theta^m_k \cos\theta^{m'}_k - \sin\theta^m_k \sin\theta^{m'}_k \\
    &= \cos(\theta^m_k + \theta^{m'}_k).
\end{align*}  

Thus, in Eq.~\eqref{eq: pauli-X}, the term $X^{mm'}_k$ is replaced
by 
\begin{equation*}
Z^{mm'}_k = \frac{\cos(\theta^m_k + \theta^{m'}_k)}{\displaystyle\cos(\theta^m_k - \theta^{m'}_k)}, 
\end{equation*}  
called the \textit{Pauli-$Z$ tensor}, as expressed in detail in 
\begin{align}\label{eq: pauli-Z}
    \expval{\sigma^z_k} = 
    \frac{\displaystyle \sum_{m,m'=1}^M c_m c_{m'}\, C^{mm'} Z^{mm'}_k}
         {\displaystyle \sum_{m,m'=1}^M c_m c_{m'}\, C^{mm'}}.
\end{align}  

This can be readily extended to multi-body operator. For instance,
for the two-body interaction $\sigma^z_k \sigma^z_{k+1}$, the
contributions to the expectation value factorize into $Z^{mm'}_k
Z^{mm'}_{k+1}$.  

These compact analytic forms make the evaluation of local
observables highly efficient, scaling only polynomially with $M$ and
$L$, which requires $O(LM^2)$ operations with $O(M^2)$ resources. 

\subsection{Gradients of the expectation value}\label{app: grad}

The SPS ansatz allows for an analytical evaluation of the derivative of the expectation value of a local observable with respect to any ansatz parameter in $\Theta$, as utilized in Secs.~\ref{sec: trainability} and~\ref{sec: GS}. In particular, the derivative of the energy with respect to an SPS parameter $\vartheta$ is given by
\begin{equation}
    \pdv{E_\mathrm{SPS}}{\vartheta}
    = \frac{1}{\mathcal{Z}}\pdv{\mel{\Psi_u}{\mathcal{H}}{\Psi_u}}{\vartheta}
    - E_\mathrm{SPS}\frac{1}{\mathcal{Z}}\pdv{\mathcal{Z}}{\vartheta}.
\end{equation}
Since the unnormalized expectation value of the Hamiltonian,
$\expval{\mathcal{H}}_u := \mel{\Psi_u}{\mathcal{H}}{\Psi_u}$,
is composed of unnormalized local Pauli expectation values, the derivative with respect to $\theta^m_l$ depends only on the site index $l$ and is independent of $m$.
As an example, for the derivative of the unnormalized local observable $\expval{\sigma^x_k}_u$ with respect to $c_m$, we have
\begin{align}
    \pdv{\expval{\sigma^x_k}_u}{c_m}
    &= 2\sum_{m'=1}^M c_{m'} \sin\qty(\theta^m_k+\theta^{m'}_k)
       \prod_{l\neq k}^L \cos\qty(\theta^m_l-\theta^{m'}_l) \nonumber\\
    &= 2\sum_{m'=1}^M c_{m'} C^{mm'} X^{mm'}_k,
\end{align}
where the factor of 2 arises from the symmetry between the indices $m$ and $m'$.

For the derivative with respect to $\theta^m_l$, two cases can be distinguished: (i) for $l = k$ and (ii) for $l \neq k$.\\
\\
(i) For $l = k$, the derivative is
\begin{align}
    &\pdv{\expval{\sigma^x_k}_u}{\theta^m_k} \nonumber\\
    &= 2\sum_{m'=1}^M c_mc_{m'}\cos\qty(\theta^m_k+\theta^{m'}_k)\prod_{l\neq k}^L \cos\qty(\theta^m_l-\theta^{m'}_l) \nonumber\\
    &= 2\sum_{m'=1}^M c_mc_{m'}C^{mm'}Z^{mm'}_k.
\end{align}
This expression maintains the same structural form as the Pauli-$Z$ tensor $Z^{mm'}_k$, which can be directly reused from the computation of the energy term $E_\mathrm{SPS}$.\\
\\
(ii) For $l \neq k$, the derivative becomes
\begin{align}
    &\pdv{\expval{\sigma^x_k}_u}{\theta^m_l} \nonumber\\ 
    &= -2\sum_{m'=1}^M c_mc_{m'}\sin\qty(\theta^m_k+\theta^{m'}_k) \nonumber\\
    &\qquad\qquad\qquad\prod_{p\neq l,k}^L \cos\qty(\theta^m_p-\theta^{m'}_p)\sin\qty(\theta^m_l-\theta^{m'}_l) \nonumber\\
    &= -2\sum_{m'=1}^M c_mc_{m'}C^{mm'}X^{mm'}_k\frac{\sin\qty(\theta^m_l-\theta^{m'}_l)}{\displaystyle\cos\qty(\theta^m_l-\theta^{m'}_l)} \nonumber\\
    & = -2\sum_{m'=1}^M c_mc_{m'}C^{mm'}X^{mm'}_k\tan\qty(\theta^m_l-\theta^{m'}_l),
\end{align}
which adds the extra tangent function beside the Pauli-$X$ tensor.

Similarly, for the local Pauli-$Z$ operator, we have
\begin{equation}
    \pdv{\expval{\sigma^z_k}_u}{\theta^m_k} = -2\sum_{m'=1}^M c_mc_{m'}C^{mm'}X^{mm'}_k
\end{equation}
and
\begin{equation}
    \pdv{\expval{\sigma^z_k}_u}{\theta^m_l} = -2\sum_{m'=1}^M c_mc_{m'}C^{mm'}Z^{mm'}_k\tan\qty(\theta^m_l-\theta^{m'}_l)
\end{equation}
for $l \neq k$. In general, for two-body terms, taking the derivative with respect to a parameter associated with one operator site transforms the corresponding Pauli tensor into another tensor of the same family.

Finally, the derivative of the normalization factor $\mathcal{Z}$ is
\begin{equation}
    \pdv{\mathcal{Z}}{\vartheta} = \begin{cases}
        2\displaystyle\sum_{m'=1}^M c_{m'}C^{mm'}: \vartheta = c_m, \\
        -2\displaystyle\sum_{m'=1}^M c_mc_{m'} C^{mm'}\tan\qty(\theta^m_l-\theta^{m'}_l): \vartheta = \theta^m_l.
    \end{cases} 
\end{equation}

As shown above, the gradient expressions with respect to the SPS parameters $\Theta$ can be implemented analytically, eliminating the need for automatic differentiation and enabling efficient parallel benchmarking during computation.

\section{Connectivity in random coupling system}\label{app: conn}

\begin{figure}
    \centering
    \includegraphics[width=1.0\linewidth]{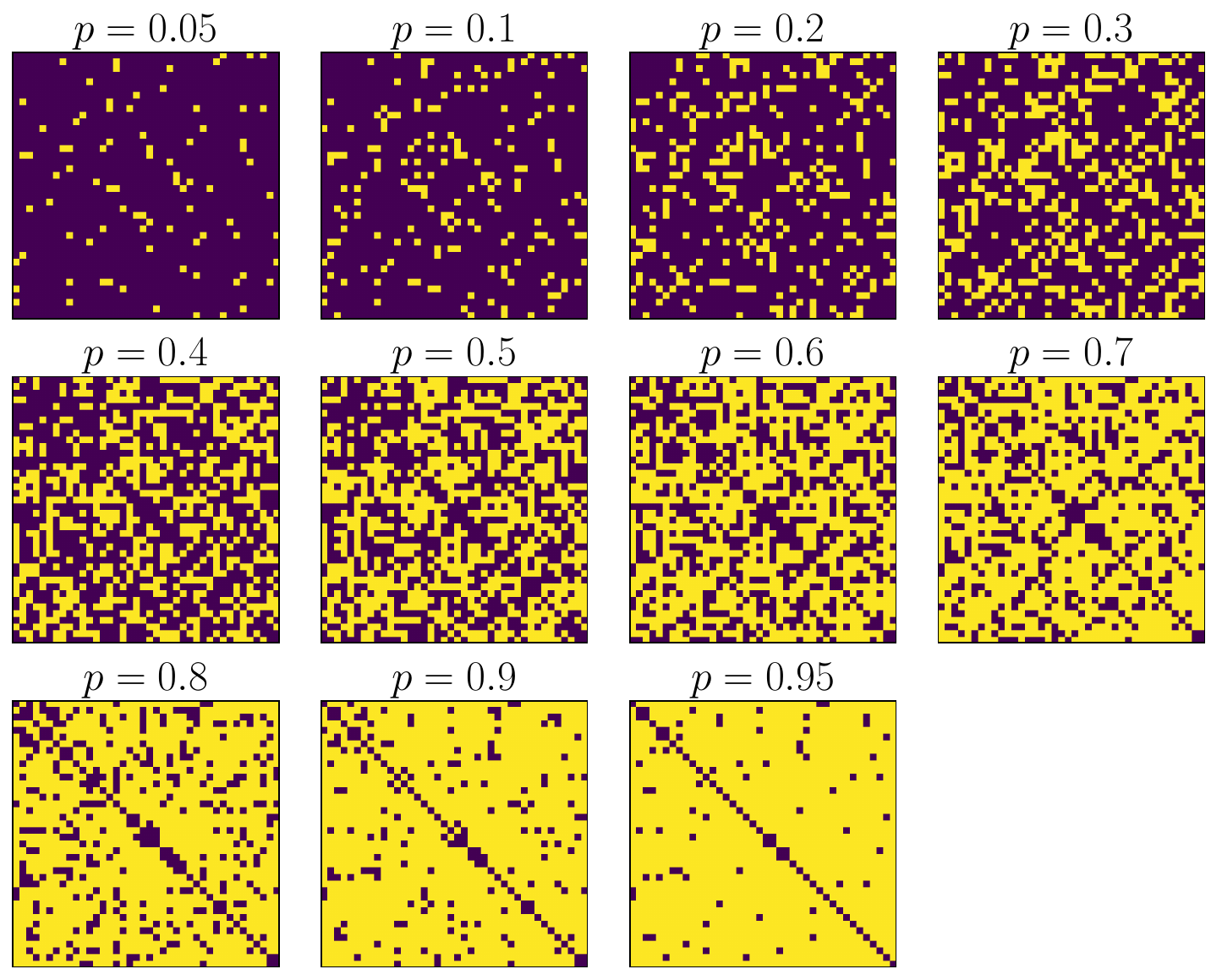}
    \caption{Adjacency matrices of randomly connected graphs in various connection probability $p$. Purple and yellow denote matrix elements with values  0 and 1, respectively.}
    \label{fig: adj_matrix}
\end{figure}

This section describes the connectivity structure of a random-coupling lattice generated with varying connection probability $p$. In the implementation, a random number generator from the NumPy package is used with a fixed seed (set to 5) to ensure reproducibility. Each possible pair of nodes $(i,j)$, with $i<j$, is connected independently with probability $p$. The resulting graph is generally equivalent to an Erd\"os-R\'enyi $G(L,p)$ network. The adjacency matrices for the different values of the probability of having connected links $p$ are pictorially given in Fig.~\ref{fig: adj_matrix}.

\end{document}